\documentclass{aa}
\usepackage[varg]{txfonts}
\usepackage[utf8]{inputenc}
\usepackage{amsmath}
\usepackage{amsfonts}
\usepackage{amssymb}
\usepackage{graphicx}
\usepackage{xcolor}

\usepackage{soul}	
\usepackage{xcolor}
\usepackage{hyperref}

\author{Rowan Batzofin \inst{\ref{inst1}}
\and Pierre Cristofari \inst{\ref{inst2}}
\and Kathrin Egberts \inst{\ref{inst1}}
}
\institute{Universit\"at Potsdam, Institut für Physik und Astronomie, Campus Golm, Haus 28, Karl-Liebknecht-Str. 24/25, 14476 Potsdam-Golm, Germany
 \label{inst1}
\and Observatoire de Paris, PSL Research University, 61 avenue de l’Observatoire, Paris, France \label{inst2}
}
\usepackage{aas_macros}
\usepackage{url}
\usepackage{hyperref}
\usepackage{natbib}
\bibpunct{(}{)}{;}{a}{}{,}
\DeclareUnicodeCharacter{03B3}{$\gamma$}
\usepackage{xcolor}
\usepackage{siunitx}

\begin{document}
\title{Young massive star clusters as teraelectronvolt emitters: Constraints from H.E.S.S. and LHAASO}
\titlerunning{Young massive star clusters as TeV emitters: Constraints from H.E.S.S. and LHAASO}

\abstract {

{Young massive star clusters (YMSCs) have been proposed as excellent candidates for the main sources of Galactic cosmic rays (CRs) up to the petaelectronvolt range. The detection and study of gamma rays in the very-high-energy (VHE, E>100 GeV) range has added to arguments in favour of this hypothesis. To date, observations with current instruments have detected only a few YMSCs. Next-generation observatories are expected to significantly increase this number, providing a larger sample that will improve our ability to constrain the role of YMSCs in the origin of CRs.}

{We studied the population of YMSCs detected in the TeV range and their properties, confronting simulations of the YMSC population to the current observed sample, to address fundamental questions regarding particle acceleration at YMSC shocks concerning the spectrum of accelerated particles, the efficiency of the CR production, and the fraction of the wind luminosity that is converted into turbulent magnetic fields.}

{We used Monte Carlo methods to simulate the Galactic population of YMSCs in the gamma--ray domain and confront our simulations to the catalogue of sources of the systematic survey of the Galactic plane performed by H.E.S.S. (HGPS) and the catalogue from the all-sky instrument LHAASO, the First LHAASO Catalogue of Gamma-Ray Sources.}

{We systematically explored the parameter space of our model, including, for example, the slope of accelerated particles, $\alpha$; the CR efficiency, $\eta_{\rm CR}$; the fraction of the wind luminosity converted into turbulent magnetic field, $\eta_{\rm b}$; and the diffusion regime. In particular, we found five possible sets of parameters for which $\gtrsim 75\%$ of Monte Carlo realisations are found to be in agreement with the combined data from the HGPS and LHAASO first catalogue: $\alpha =4.5$, $\eta_{\rm CR} = 10^{-2.0}$, and $\eta_{\rm b} = 10^{-2.0}$, $L_{\rm inj} = 1$ pc---assuming the Kraichnan diffusion regime---and $\alpha = 4.4$, $\eta_{\rm CR} = 10^{-2.5}$, $\eta_b = 10^{-3.0}$, with $L_{\rm inj}  = 0.1$ pc, assuming the Kolmogorov diffusion regime. Certain scenarios and regions of parameter space are strongly disfavoured, such as the Bohm diffusion regime at YMSCs.}

{Our model successfully reproduces the YMSC population observed in both the HGPS and the First LHAASO Catalogue of Gamma-Ray Sources. With future systematic surveys, such as those by the Cherenkov Telescope Array Observatory (CTAO), this approach will help break degeneracies and improve our understanding of particle acceleration at YMSC shocks in the Galaxy.
}
}


\maketitle
\nolinenumbers
\section{Introduction}
Cosmic rays (CRs) were discovered thanks to the pioneering work of Domenico Pacini and Victor Hess in the 1910s and have been extensively studied ever since  \citep{Hess:1912srp}. Remarkably, the most fundamental question of CR physics remains unanswered, that is, we do not know where they are produced~\citep{blasi2013,gabici2019}. Among the leading hypotheses is that they are accelerated by young massive star clusters (YMSCs;~\citealp{aharonian2019,Morlino_2021,gabici2023,Menchiari_2023,Padilha_2025}), via the first-order Fermi mechanism called diffusive shock acceleration (DSA;~\citealp{axford1977,krymskii1977,Blandford_1978, bell1978}). DSA can efficiently energise protons, ions, and electrons. Subsequently, the accelerated protons and electrons can interact with the interstellar medium (ISM) to produce gamma rays in the very-high-energy (VHE) domain, mainly through two mechanisms: 1) the production and decay of neutral pions in the collision of accelerated protons with hadrons of the ISM (hadronic mechanism); and 2) the inverse Compton scattering of accelerated electrons on soft photons: cosmic microwave background (CMB), infrared, and optical (leptonic mechanism). Given the high densities and magnetic fields inferred in YMSCs the hadronic process is thought to be dominant.

 In particular, the High Energy Stereoscopic System (H.E.S.S.), an array of imaging atmospheric Cherenkov telescopes located in the Khomas highlands in Namibia, performed a systematic survey of the Galactic plane. The H.E.S.S. Galactic Plane Survey (HGPS) covers the Galactic plane from longitudes of $250^{\circ}$ to $65^{\circ}$ and latitudes of $-3^{\circ}$ to $3^{\circ}$. This systematic survey found 78 sources in the TeV domain. Among these sources is the stellar cluster Westerlund 1 \citep{westerlund_1}. LHAASO (Large High Altitude Air Shower Observatory), a multi-purpose ground-based extensive air shower array located in Daocheng County, Sichuan Province, China, released a catalogue of very-high-energy and ultra-high-energy gamma-ray sources \citep{LHAASO_catalogue}. The First LHAASO Catalogue of Gamma-Ray Sources covers the sky above 1 TeV, in declinations from $-20^{\circ}$ to $80^{\circ}$. The catalogue contains 90 sources, 43 of which show significant detections above 100 TeV. Among the gamma-ray emissions observed by LHAASO, some originate from the superbubble in the direction of the Cygnus X star-forming region \citep{LHAASO_cygnus}.

In this work, we intend to simulate the population of TeV-emitting YMSCs and confront it with the population detected in the HGPS and the First LHAASO Catalogue of Gamma-Ray Sources. For this purpose, we simulated populations of Galactic YMSCs using a Monte Carlo method, in an approach inspired by what was done for the supernova remnant (SNR) population~\citep{cristofari2013,1st_paper}. We simulated the positions, masses, and wind luminosities of the YMSCs and calculated the gamma-ray emission from the YMSCs at the current time. We constrained the simulations with measurements from the HGPS and the First LHAASO Catalogue of Gamma-Ray Sources by identifying the detectable sources in the simulations and comparing them with the detected sample of YMSCs. For the HGPS the detectability is based on the longitude, latitude, and angular-extension-dependent HGPS sensitivity, which is visualised in the orange area of Fig.~\ref{fig:single_realisation} for a point-source luminosity of $10^{34}$~photons/s. The inclusion of the complex dependencies of the HGPS sensitivity is crucial for drawing conclusions on the YMSC population based on this data comparison. For the LHAASO detectability we considered the two instruments---Water Cherenkov Detector Array (WCDA) and Kilometre Squared Array (KM2A)---separately and took into account the declination and extent of the simulated sources, which is visualised in the blue-green area of Fig.~\ref{fig:single_realisation} for a point-source luminosity of $10^{33}$~photons/s and $3\times 10^{28}$~photons/s for the WCDA and KM2A, respectively.
\paragraph{}
The paper is structured as follows. In Sect.~\ref{section:pop_model}, we introduce the population model, detailing the physics of individual YMSCs and the distributions adopted to place them within the Milky Way. Sect.~\ref{section:simulating_populations} describes the methods used to compare our simulated populations with the HGPS. We present our results in Sect.~\ref{section:results}, and conclude in Sect.~\ref{section:conclusion}.

\section{YMSC population model} 
\label{section:pop_model}

Following the approach of other population simulations~\citep{1st_paper}, simulating the Galactic population of YMSCs emitting in the TeV range requires: (a) a computationally efficient model for particle acceleration at the YMSC and the resulting gamma-ray emission from hadronic interactions with the surrounding medium, and (b) a description of the spatial distribution of YMSCs in the Galaxy.

\subsection{Modelling the gamma-ray emission of YMSCs} \label{section:gamma-ray_calc}
We followed the approach of \citet{Menchiari_2023} and \citet{Menchiari_2024} for the modelling of the gamma-ray emission produced by YMSCs. We assume that the collective wind of a YMSC can be described in the same way as the wind of a single star with the total mass of the cluster in the centre, the mass-loss rate as the sum of all the mass-loss rates in the cluster, and the wind luminosity as the sum of all the wind luminosities in the cluster.

\subsubsection{Dynamics of the YMSC winds}
\label{section:ymsc_evolution}
Following the work of \citet{Weaver_1977} the stellar wind is described by three distinct regions: an inner region filled with cold supersonic wind from the central source or cluster core, a cavity of hot shocked wind material, and an outer shell composed of the swept-up ISM. The inner wind region and the cavity are separated by a termination shock (TS), which is located at radius $R_{\rm TS}$. There is a contact discontinuity between them, separating the cavity from the swept-up shell at radius $R_{\rm CD}$. The forward shock is at radius $R_{\rm FS}$. 
\paragraph{}
The evolution of the cluster's wind-blown bubble has three stages, depending on the onset of the radiative losses. The typical cooling timescale is \citep{Draine_2011} $t_{\rm cool} \simeq 60 \left(\frac{n}{0.001 \rm \: cm^{-3}}\right)^{-1} \left(\frac{T}{10^6 \rm \: K}\right)^{1.7} \: \rm Myr$, where n is the density and $T$ is the temperature, both of which are normalised to typical values of the shocked wind and the numerical coefficient refers to Solar metallicity. Eventually, radiative losses lead to the collapse of the outer shell, and we can assume $R_{\rm CD} \simeq R_{\rm FS} \equiv R_{\rm b}$.
\paragraph{}
Since we were only interested in YMSCs, a simple approximation of the bubble size was obtained by considering the momentum of the material in the thin shell between $R_{\rm CD}$ and $R_{\rm FS}$. This momentum evolves due to the work done by the pressure inside the hot bubble and the change in the bubble’s internal energy. Using this, the bubble radius can then be estimated as follows~\citep{Morlino_2021}:

\begin{equation}
    R_{\rm b}\approx 99 \left(\frac{L_{\rm w}}{2\times 10^{38} \rm erg \: s^{-1}}\right)^{1/5} \left(\frac{\rho_{0}}{10m_{\rm p} \rm \: cm^{-3}}\right)^{-1/5} \left(\frac{t_{\rm age}}{3 \rm Myr}\right)^{3/5} \rm pc 
    \label{equation:Rb}
    ,
\end{equation}
where $\rho_{\rm 0}$ is the density of the local ambient medium in which the bubble expands (in this work we assumed this to be $10 m_p \: \rm cm^{-3}$, where $m_p$ is the mass of a proton), $t_{\rm age}$ is the age of the YMSC, and $L_{\rm w}$ is the wind luminosity. The TS position is the position at which the ram pressure of the wind and the pressure of the hot gas in the cavity are equal:
\begin{multline}
    \label{equation:Rts}
    R_{\rm TS} \approx 17 \left(\frac{\dot{M}}{10^{-4}\rm M_{\odot} \: yr^{-1}}\right)^{1/2} \left(\frac{v_{\rm w}}{2500 \rm km \: s^{-1}}\right)^{1/2} \\ \left(\frac{L_{\rm w}}{2\times 10^{38} \rm erg \: s^{-1}}\right)^{-1/5}  \left(\frac{\rho_{0}}{10m_{\rm p} \rm \: cm^{-3}}\right)^{-3/10} \left(\frac{t_{\rm age}}{3 \rm Myr}\right)^{2/5} \rm pc
    ,
\end{multline}
where $\dot{M}$ is the mass-loss rate of the wind and $v_{\rm w} = \sqrt{\frac{2 L_{\rm w}}{\dot{M}}}$ is the velocity of the wind.

\subsubsection{Particle acceleration at YMSC winds}
\label{section:CR_acc}
We only considered the total flux of the cluster and did not look at the individual sources in the cluster. Particles are accelerated at the TS via the diffusive shock acceleration process. Assuming stationarity and a radial symmetry, the distribution of CRs, $f(r, p)$, can be found by solving the steady-state transport equation in 1D spherical symmetry~\citep{Morlino_2021}:
\begin{equation}
    \frac{\partial}{\partial r} \left[r^2 D(r,p) \frac{\partial f}{\partial r}\right] - r^2 u(r) \frac{\partial f}{\partial r} + \frac{d[r^2 u(r)]}{dr} \frac{p}{3} \frac{\partial f}{\partial p} + r^2 Q(r,p) = 0
    \label{equation:transport_equation}
    ,
\end{equation}
where $p$ is the particle momentum, $u(r)$ is the plasma speed and $D(r,p)$ is the diffusion coefficient. The source term $Q(r,p)$ describes the particle injection taking place at the TS; this is described by:
\begin{equation}
    Q(r,p) = \frac{\eta_{\rm inj} n_{\rm w} \nu_{\rm w}}{4 \pi p_{\rm inj}^2} \delta (p-p_{\rm inj}) \delta (r-R_{\rm TS}),
    \label{equation:Q}
\end{equation}
where $n_{\rm w}$ is the density upstream of the shock, $\nu_w$ is the speed of the wind, $p_{\rm inj}$ is the momentum of the injected particles, and $\eta_{\rm inj}$ is the fraction of particles that are injected into the acceleration process. Equation~\eqref{equation:transport_equation} is solved in three zones ($r<R_{\rm TS}$, $R_{\rm TS} < r < R_{\rm b}$, $r > R_{\rm b}$):
\begin{equation}
\label{equation:proton_spectrum}
    f = \begin{cases}
    f_{\rm TS}(p) \: {\rm exp}\left[- \int_r^{R_{\rm TS}} \frac{v_w}{D_w(r',p)}dr' \right] & r \leq R_{\rm TS} \\
    f_{\rm TS}(p)e^\alpha \frac{1+\beta(e^{\alpha_{b} \; - \alpha} -1)}{1+\beta(e^{\alpha_{b}} -1)} & R_{\rm TS} \leq r \leq R_{\rm b} \\
    f_{\rm TS}(p) \frac{e^{\alpha_{b}}}{1+\beta (e^{\alpha_{b}} -1)} \frac{R_{\rm b}}{r} & r \geq R_{\rm b}
    \end{cases},
\end{equation}
where 
\begin{equation}
    \alpha(r,p) = \frac{v_{\rm b} R_{\rm TS}}{D_b(p)}\left(1- \frac{R_{\rm TS}}{r}\right),
\end{equation}
\begin{equation}
    \alpha_b = \alpha(r=R_b, p)
\end{equation}
and
\begin{equation}
    \beta(p) = \frac{D_{\rm ism}(p) R_b}{v_b R^2_{\rm TS}},
\end{equation}
where $v_{\rm b} = v_w/4$ is the speed immediately downstream of the shock. $D_w$, $D_{\rm b}$, and $D_{\rm ism}$ are the diffusion coefficients in the wind region, in the cavity, and outside the bubble, respectively. The distribution at the TS, $f_{\rm TS}$, is described by
\begin{equation}
    f_{\rm TS}(p) \simeq \frac{3 n_w v_w^2 \eta_{\rm CR}}{4 \pi \Lambda_p \left(m_pc\right)^3 c^2}\left(\frac{p}{m_p c}\right)^{-s}  e^{-\Gamma(p)},
\end{equation}
where c is the speed of light, $\eta_{\rm CR}$ is the fraction of the wind luminosity converted into CR luminosity, and $\Lambda_p$ is the normalisation factor
\begin{equation}
    \Lambda_p = \int_{\rm x_{\rm inj}}^\infty x^{2-s} e^{-\Gamma(x)}\left(\sqrt{1+x^2}-1\right)dx,
\end{equation}
with $x = p/(m_p c)$. The function $\Gamma(p)$ contains information about the spherical geometry and diffusion coefficients in both the wind region and the hot cavity. A good approximation for $e^{\Gamma(p)}$ is \citep{Morlino_2021}
\begin{equation}
    \label{equation:diffusion_coefficient}    e^{\Gamma (p)} \simeq \left[1+a_1 \left(\frac{p}{p_{\rm max}}\right)^{a_2}\right]e^{-a_3 (p/p_{\rm max})^{a_4}},
\end{equation}
where $p_{\rm max}$ is the upstream maximum momentum defined by the condition that the upstream diffusion length is equal to the shock radius, i.e. $D_w (p_{\rm max}) = v_w R_{\rm TS}$. $a_1$, $a_2$, $a_3$, and $a_4$ are parameters that mainly depend on the adopted diffusion model, we used the same parameters as in \citet{Menchiari_2024}, which are summarised in Table~\ref{table:diffusion_params}.
\begin{table}[h]
    \centering
    \caption{Parameter values used to calculate the distribution function in Equation~\ref{equation:diffusion_coefficient} for different diffusion regimes.}
    \begin{tabular}{|c|c|c|c|c|}
	\hline
	Models & $a_1$ & $a_2$ & $a_3$ & $a_4$ \\
	\hline
	Kolmogorov & 10 & 0.308653 & 22.0241 & 0.43112 \\
	Kraichnan & 5 & 0.448549 & 12.52 & 0.642666 \\
	Bohm & 8.94 & 1.29597 & 5.31019 & 1.13245 \\
	\hline

	\end{tabular}
    
    \label{table:diffusion_params}
\end{table}

\subsubsection{Maximum energy of accelerated particles}
The maximum energy of particles can be estimated using only the intrinsic properties of the YMSC by equating the particle diffusion length to the size of the TS:
\begin{equation}
    \label{equation:emax_gen}
    \frac{D_w (E_{\rm max})}{\nu_w}=R_{\rm TS}.
\end{equation}
We considered three different diffusion regimes: Kolmogorov, Kraichnan, and Bohm. Since the turbulence is most likely generated by the wind itself \citep{Blasi_2023}, we assume that some fraction, $\eta_B$, of the wind luminosity upstream is converted into magnetic flux:
\begin{equation}
    \label{equation:frac_wind_luminosity}
    4\pi r^2v_{\rm w} \frac{\delta B_{\rm w}^2}{4\pi}=\eta_B \frac{1}{2}\dot{M}v_{\rm w}^2.
\end{equation}
At the TS we assume that the magnetic field is compressed by the shock and forms a strong shock, such that we have $\delta B_{\rm b} = \sqrt{11}\delta B_{\rm w}(R_{\rm TS})$. The diffusion coefficients are then
\begin{equation}
    \label{equation:D_kol}
    D_{\rm Kol} = \frac{1}{3}v_p r_L^{1/3} L_{\rm inj}^{2/3},
\end{equation}

\begin{equation}
    \label{equation:D_kra}
    D_{\rm Kra} = \frac{1}{3}v_p r_L^{1/2} L_{\rm inj}^{1/2},
\end{equation} and
\begin{equation}
    \label{equation:D_bohm}
    D_{\rm Bohm} = \frac{1}{3}v_p r_L,
\end{equation}
where $v_{\rm p}$ is the particle velocity, $r_{\rm L} = \frac{c p}{e\,\delta B_{\rm inj}}$ is the particle Larmor radius, $L_{\rm inj}$ is the turbulence injection scale, and $\delta B_{\rm inj} = \delta B(L_{\rm inj})$. The maximum energy is naturally influenced by the diffusion regime. The maximum energy for each of the three diffusion regimes is given by \citep{Morlino_2021}

\begin{multline}
    \label{equation:emax_kol}
    E_{\rm max}^{\rm Kol} = 1.2 \left( \frac{\eta_B}{0.1} \right)^{1/2} \left(\frac{\dot{M}}{10^{-4} \: M_{\rm \odot}yr^{-1}} \right)^{-3/4} \left( \frac{L_{\rm w}}{10^{39} \: \rm erg \: s^{-1}} \right)^{37/20} \\ \left( \frac{\rho_0}{20 \: m_p \: \rm cm^{-3}} \right)^{-3/5} \left( \frac{t_{\rm age}}{3 \: \rm Myr} \right)^{4/5} \left(\frac{L_{\rm inj}}{2 \: \rm pc} \right)^{-2} \rm PeV,
\end{multline}

\begin{multline}
    \label{equation:emax_kra}
    E_{\rm max}^{\rm Kra} = 2.8 \left( \frac{\eta_B}{0.1} \right)^{1/2} \left(\frac{\dot{M}}{10^{-4} \: M_{\rm \odot}yr^{-1}} \right)^{-5/10} \left( \frac{L_{\rm w}}{10^{39} \: \rm erg \: s^{-1}} \right)^{13/10} \\ \left( \frac{\rho_0}{20 \: m_p \: \rm cm^{-3}} \right)^{-3/10} \left( \frac{t_{\rm age}}{3 \: \rm Myr} \right)^{2/5} \left(\frac{L_{\rm inj}}{2 \: \rm pc} \right)^{-1} \rm PeV,
\end{multline} and

\begin{equation}
    \label{equation:emax_bohm}
    E_{\rm max}^{\rm Bohm} = 10 \left( \frac{\eta_B}{0.1} \right)^{1/2} \left(\frac{\dot{M}}{10^{-4} \: M_{\rm \odot}yr^{-1}} \right)^{-1/4} \left( \frac{L_{\rm w}}{10^{39} \: \rm erg \: s^{-1}} \right)^{3/4} \rm PeV.
\end{equation}

\subsubsection{Gamma rays}
Gamma rays from stellar clusters are produced through the decay of neutral pions generated in inelastic proton--proton collisions. The spectrum and morphology depend on the distribution of cosmic rays and on the properties of the target medium. In this work, we made the simplifying assumption that the target medium has the same density as the ambient material in which the bubble expands (set to $10\,m_{\rm p}\,{\rm cm}^{-3}$ in this work). The gamma-ray emission from accelerated protons was computed using the {\sc Naima} Python package \citep{naima}. The hadronic emission was calculated using the pion-decay model \citep{naima_PD}, and the proton spectrum is described by  Equation~\eqref{equation:proton_spectrum}. We assume that only the hadronic component contributes to the gamma-ray emission.

\subsection{Distribution of Galactic YMSCs}

\label{section:source_distribution}

\subsubsection{Distribution of the physical parameters}

\citet{Hunt_2023} completed an open cluster catalogue using the latest data release, DR3 \citep{Gaia_DR3}, from the {\it{Gaia}} satellite \citep{Gaia_2016}. They catalogued 7167 clusters; however, with more stringent cuts the catalogue contains 4105 highly reliable clusters. \citet{Hunt_2024} used their catalogue to determine the distribution of the ages and masses of clusters in the Milky Way. They derived completeness-corrected age and mass functions. The catalogue is only complete within 1.8 kpc of the Sun and for objects heavier than $230$ $M_{\rm \odot}$. The age function of clusters is given by a broken power law as
\begin{equation}
    n(t) \propto \begin{cases}
    \left(t/{t_{\rm break}}\right)^{-0.594} & \rm for \: t < t_{\rm break} \\
    \left(t/{t_{\rm break}}\right)^{-2.321} & \rm for \: t \geq t_{\rm break} \\ \end{cases},
    \label{equation:cluster_formation_rate}
\end{equation}
with $t_{\rm break} = 10^{8.33 \pm 0.04}$yr, $n(t)$ in units of clusters per parsec squared per year. 
The mass function is well approximated by a power law with index $\kappa = -2$. However, \citet{Hunt_2024} provided different mass distributions for different cluster age ranges, each of which follows a power law with a distinct index. In this work, we primarily used the mass function corresponding to clusters younger than $10 \; {\rm Myr}$:

\begin{equation}
    \label{equation:cluster_mass_func}
    f(M) \propto M^{-1.87},
\end{equation}
where $f(M)$ is in units of clusters/$\text{pc}^2$/M$_{\odot}$/Myr. 
\paragraph{}
Each cluster is composed of numerous stars and these stars also have a mass distribution. In this work we adopted the following expression for the initial mass distribution in the cluster \citep{kroupa_mnras_322_2001}:
\begin{equation}
    \label{equation:star_mass_dist}
    f_{\rm \star}(M_{\rm \star}) \propto \frac{dN_{\rm \star}}{dM_{\rm \star}} = \begin{cases}
    M_{\rm \star}^{-0.3} & {\rm for }\: M_{\rm \star} < 0.08 M_{\rm \odot}\\
    0.08 M_{\rm \star}^{-1.3} & {\rm for }\: 0.08 M_{\rm \odot} \leq M_{\rm \star} \leq 0.5 M_{\rm \odot}\\
    0.04 M_{\rm \star}^{-2.3} & {\rm for }\: M_{\rm \star} > 0.5 M_{\rm \odot}
    \end{cases}
\end{equation}
With this, we can also determine the expected number of stars in a cluster of a given mass using the following equation:
\begin{equation}
    \label{equation:no_stars}
    N_{\rm \star}(M) = M \frac{\int_{\rm M_{\rm \star,\: \rm min}}^{M_{\rm \star,\: \rm max}} f_{\rm \star}(M_{\rm \star})dM_{\rm \star}}{\int_{\rm M_{\rm \star,\: \rm min}}^{M_{\rm \star,\: \rm max}} M_{\rm \star} f_{\rm \star}(M_{\rm \star})dM_{\rm \star}},
\end{equation}
where $M_{\rm \star,\: \rm min}$ and $M_{\rm \star,\: \rm max}$ are the minimum and maximum stellar masses that can be generated in a cluster, respectively. The value of $M_{\rm \star,\: \rm min}$ is $0.08M_{\rm \odot}$, which is the minimum theoretical mass to support significant nuclear burning \citep{Carrol_1996}. Following this mass distribution we find that our clusters have $\sim 4\%$ OB stars. The radius and luminosity of each star can be estimated from its mass using a combination of mass--luminosity relationships described by \citet{Eker_2018} and \citet{Yungelson_2008}:
\begin{equation} \label{equation:stellar_masses}
    L_{\rm \star} = \begin{cases}
    L_{\rm b1}\left(\frac{M_\star}{M_{\rm b1}}\right)^{\alpha_1} \left[\frac{1}{2}+\frac{1}{2}\left(\frac{M_{\rm \star}}{M_{\rm b1}}\right)^{1/\triangle_1}\right]^{(-\alpha_1 + \alpha_2)\triangle_1} & \rm for \: \frac{M_\star}{M_{\rm \odot}} \leq 12\\ 

    \kappa L_{\rm b2}\left(\frac{M_\star}{M_{\rm b2}}\right)^{\alpha_2} \left[\frac{1}{2}+\frac{1}{2}\left(\frac{M_{\rm \star}}{M_{\rm b2}}\right)^{1/\triangle_2}\right]^{(-\alpha_2 + \alpha_3)\triangle_2} & \rm for \: \frac{M_\star}{M_{\rm \odot}} \geq 12
    \end{cases},
\end{equation}
where $L_{\rm b1} = 3191 \: L_\odot$ and $L_{\rm b2} = 368874 \: L_\odot$ are the luminosity values at the mass break points $M_{\rm b1} = 7 \: M_\odot$ and $M_{\rm b2} = 36.089 \: M_\odot$, respectively. The power law indices are $\alpha_1 = 3.97, \: \alpha_2 = 2.86,$ and $\alpha_3 = 1.34$. To smooth the function across the different power-law components, $\triangle_1$ and $\triangle_2$ were set to 0.01 and 0.15, respectively. $\kappa$ is the normalisation factor, so there is continuity at 12 $M_\odot$ equal to 0.817. The radius of a star can be estimated using the mass--radius relation from \citet{Demircan_1991}:
\begin{equation}
    R_\star = 0.85 \left(\frac{M_\star}{M_\odot}\right)^{0.67} R_\odot.
\end{equation}

The mass-loss rate of each star in the star cluster can be estimated using the mass of the star \citep{Nieuwenhuijzen_1990}:
\begin{equation}
\begin{aligned}
    \label{equation:mdot_luminosity}
    \rm log \left(\frac{\dot{M}_\star}{M_\odot yr^{-1}}\right) = -14.2 + 1.24 log\left(\frac{L_\star}{L_\odot}\right) \\ \rm + 0.16log\left(\frac{M_\star}{M_\odot}\right) + 0.81\left(\frac{R_\star}{R_\odot}\right).
\end{aligned}
\end{equation}
where the wind luminosity of a star can be estimated from the terminal wind velocity, $v_{\inf}$:
\begin{equation}
    L_{\rm w, \: \star} = \frac{1}{2}\dot{M}_\star v_{\rm \inf}^2
\end{equation}
 $v_{\inf}$ is defined as \citep{Kudritzki_2000}
\begin{equation}
    v_{\inf} = C(T_{\rm eff}) \left[\frac{2GM_\star(1-\Gamma)}{R_\star}\right]^{1/2},
\end{equation}
where G is the gravitational constant and $C(T_{\rm eff})$ is a parameter dependant on the star temperature, $T_{\rm eff}$, which is 2.65 for $T_{\rm eff} > 21000 \: \rm K,$ $1.4$ for $10000$ K $< T_{\rm eff} < 21000$ K, and 1.0 for $T_{\rm eff} \leq 10000$ K \citep{Kudritzki_2000}. 
\subsubsection{Spatial distribution of YMSCs}
\begin{table*}[h]
\caption{Parameter values for arms in the Reid distribution of sources.}
\label{table:Reid_arms}
\centering
\small
\begin{tabular}{|c|c|c|c|c c|}
\hline
Spiral arm & $\beta$ Range (deg) & $\beta_{\text{kink}}$ (deg) & $R_{kink}$ (kpc) & \multicolumn{2}{|c|}{$\psi$ (deg)} \\
 & & & & $(\beta<\beta_{\text{kink}})$ & $(\beta>\beta_{\text{kink}})$ \\
\hline
Sagittarius-Carina & $2 \rightarrow 97$ & 24 & 6.04 & 17.1 & 1.09 \\
Scutum-Centaurus & $0 \rightarrow 104$ & 23 & 4.91 & 14.1 & 12.1\\
Perseus & $-23 \rightarrow 115$ & 40 & 8.87 & 10.3 & 8.7 \\
Norma & $5 \rightarrow 54$ & 18 & 4.46 & -1.0 & 19.5 \\
Local & $-8 \rightarrow 34$ & 9 & 8.26 & 11.4 & 11.4 \\
3 kpc & $15 \rightarrow 18$ & 15 & 3.52 & -4.2 & -4.2 \\
Outer & $-16 \rightarrow 71$ & 18 & 12.24 & 3.0 & 9.4 \\
\hline

\end{tabular}
\end{table*}

For the distribution of YMSCs we used a four-spiral-arm model based on data from massive stars with maser parallaxes described in \citet{Reid_2019}. This model also includes some additional features: the Local Arm; a 3~kpc arm around the centre of the Galaxy; and an outer arm, which connects with the Norma Arm. The Local Arm is described as an arm segment and does not link with any of the other arms. The 3~kpc arm is associated with the Galactic bar and might not be a true spiral arm. The spiral arms are described by Equation~\ref{equation:Reid_arms} and Table~\ref{table:Reid_arms}:
\begin{equation}
    \text{ln}\left(\frac{R}{R_{\rm kink}}\right) = -\left(\beta - \beta_{\rm kink}\right) \text{tan}\psi \, ,
    \label{equation:Reid_arms}
\end{equation}
where $R$ is the galactocentric radius at azimuth $\beta$, $R_{\rm kink}$ is the radius of the kink in the arm at azimuth $\beta_{\rm kink}$, and $\psi$ is the pitch angle. In addition to this model we also included a Galactic bar described in \citet{Sormani_2022}, where they created an analytical model by fitting a four-component bar (three barred components and an axisymmetric disc) to density and kinematics observational data of the inner Galaxy. The total density is $\rho(x, y, z) = \rho_{\rm bar, 1} + \rho_{\rm bar, 2} + \rho_{\rm bar, 3} + \rho_{\rm disc}$, where the first two components are the bar and bulge, the third component is the vertically flat extension of the bar and the disc is an axisymmetric component that covers the region outside the bar. The densities of each component are as follows:
\begin{equation}
    \rho_{\rm bar, 1} (x, y, z) = \rho_1 {\rm sech}(a^m)\left[1 + \alpha (e^{-a_+^n} +e^{-a_-^n})\right]e^{-\left(\frac{r}{r_{\rm cut}}\right)^2},
\end{equation}
where
\begin{equation}
    a = \left( \left[ \left( \frac{|x|}{x_1} \right)^{c_{\perp}} + \left(\frac{|y|}{y_1}\right)^{c_{\perp}}\right]^{\frac{c_{\parallel}}{c_{\perp}}} + \left(\frac{|z|}{z_1}\right)^{c_\parallel}\right)^{\frac{1}{c_\parallel}};
\end{equation}

\begin{equation}
    a_{\pm} = \left[ \left( \frac{x \pm cz}{x_c} \right)^{2} + \left(\frac{y}{y_c}\right)^{2}\right]^{\frac{1}{2}};
\end{equation}

\begin{equation}
    r = (x^2 + y^2 + z^2)^{\frac{1}{2}},
\end{equation}
where $\alpha$ quantifies the strength of the X-shape and c quantifies its slope in the $(x, z)$ plane;

\begin{equation}
    \rho_{\rm bar, i} (x, y, z) = \rho_i e^{-a_i^{n_i}} {\rm sech}^2 \left(\frac{z}{z_i}\right) e^{-\left(\frac{R}{R_{i, \rm out}}\right)^{n_{i \rm , out}}} e^{-\left(\frac{R_{i \rm, in}}{R}\right)^{n_{i \rm , in}}},
\end{equation}
where i = {2,3}; and
\begin{equation}
    a_{i} = \left[ \left( \frac{|x|}{x_i} \right)^{c_{\perp , i}} + \left(\frac{|y|}{y_i}\right)^{c_{\perp , i}}\right]^{\frac{1}{c_{\perp ,i}}},
\end{equation}

\begin{equation}
    R = (x^2 + y^2)^{\frac{1}{2}},
\end{equation}

\begin{equation}
    \rho_{\rm disc}(R,z) = \frac{\sum_0}{4z_{\rm d}}e^{-\left(\frac{R}{R_{\rm d}}\right)^{n_{\rm d}}} e^{-\frac{R_{\rm cut}}{R}} {\rm sech} \left(\frac{|z|}{z_{\rm d}}\right)^{m_{\rm d}}.
\end{equation}

The best-fit parameters can be found in \citet{Sormani_2022}, Table 1.

\subsubsection{Generating YMSCs}
The simulated YMSCs were generated using a Monte Carlo approach. The age of each cluster was drawn from the distribution given in Equation~\ref{equation:cluster_formation_rate}, and the total mass of each YMSC was sampled from the distribution in Equation~\ref{equation:cluster_mass_func}. The number of stars in a cluster was then determined from its mass using Equation~\ref{equation:no_stars}. A mass for each star was drawn from the distribution in Equation~\ref{equation:star_mass_dist}, from which the stellar wind luminosity and mass-loss rate were estimated. The total wind luminosity of the cluster was obtained by summing the individual contributions of all stars. The clusters were placed in the Milky Way according to the model described in Sect.~\ref{section:source_distribution}. In this study, we were only interested in YMSCs, which we define as clusters that are younger than $10\,{\rm Myr}$ and have a total mass greater than $1000\,M_\odot$. The total number of such clusters can be estimated using

\begin{equation}
    \label{equation:no_clusters}
    N_{\rm SC} = D \int_{\rm M_{\rm min}}^{M_{\rm max}} \int_{\rm t_{\rm min}}^{t_{\rm max}} \int_{\rm 0}^{R_{\rm MW}} r \xi_{\rm SC} \left(M,t,r\right)dM \: dt \: dr \: d\theta,
\end{equation}
where $R_{\rm MW}$ is the radius of the Milky Way, $\xi_{\rm SC}$ is the cluster distribution function, and $D$ is a normalisation constant. We used $\xi_{\rm SC}$ in the following form:
\begin{equation}
    \label{equation:cluster_dist_func}
    \xi_{\rm SC}(M,t,r) = f(M)\,\nu(t)\,\rho(r,\theta),
\end{equation}

where $f(M)$ is the cluster initial mass function, $\nu(t)$ is the cluster formation rate, and $\rho(r,\theta)$ is the spatial distribution. We used Equation~\eqref{equation:cluster_mass_func} to approximate $f(M)$ and $\nu(t)$, and the spatial distribution described in Sect.~\ref{section:source_distribution} for $\rho(r,\theta)$.

To find the normalisation constant, we considered all the clusters in the completeness region (within 1.8 kpc) of the open cluster census (2894 clusters). We calculated the number of expected clusters of all ages and masses within 1.8 kpc of the Sun according to Equation~\eqref{equation:no_clusters} and solved for D such that the number of clusters was 2894. When calculating the normalisation constant, D, we used all the mass functions provided in \citet{Hunt_2024} rather than just that for clusters of <10 Myr since we normalised to the all the clusters in the completeness limit. We were then able to estimate the number of YMSCs we expected in our population by setting $M_{\rm min} = 1000M_\odot$, $M_{\rm max} = 6.5\times 10^{4} M_\odot$, $t_{\rm min} = 0$, $t_{\rm max} = 10$ Myr, and $R_{\rm MW} = 15$ kpc. This gave us an estimate of $\sim 294$ YMSCs.

\section{Confronting our simulated populations to available experimental data} \label{section:simulating_populations}

Following the modelling described in Sect.~\ref{section:pop_model}, we were able to simulate the Galactic population of YMSCs. The main physical quantities of interest are: 
\begin{itemize}
    \item $\alpha$, the spectral index of the accelerated particles
    \item $\eta_{\rm CR}$, the fraction of the wind luminosity converted into CRs
    \item $\eta_b$, the fraction of the wind luminosity converted into turbulent magnetic fields
    \item $L_{\rm inj}$, the turbulence injection scale
    \item the diffusion regime
\end{itemize}

We computed 100 realisations for a given set of parameters to ensure the stability of our results. Note that the positions of the simulated YMSCs were re-drawn for each of the 100 realisations for a given parameter set, but these 100 position sets are kept fixed when varying the population properties. This approach allowed us to isolate and examine the impact of each parameter on the resulting population. 

The explored parameter ranges are physically motivated and bracket the values commonly discussed in the literature. The spectral index $4.0 \leq \alpha \leq 4.5$ encompasses the canonical strong-shock prediction $\alpha \simeq 4$ from test-particle DSA \citep{axford1977,bell1978} as well as the steeper spectra ($\alpha \sim 4.1$--$4.4$) inferred in realistic Galactic CR source models and YMSC environments \citep{strong2007,gabici2019}. The CR efficiency range $10^{-3} \leq \eta_{\rm CR} \leq 10^{-1}$ spans inefficient acceleration up to the canonical $\sim10\%$ efficiency often invoked for CR sources \citep{blasi2013}, while allowing for lower efficiencies expected in distributed cluster shocks \citep{Morlino_2021}. The magnetic turbulence fraction $10^{-3} \leq \eta_b \leq 10^{-1}$ covers weak to moderately amplified magnetic fields, which is consistent with theoretical expectations of shock-driven amplification \citep{Blasi_2023}. The injection scale $0.1\,{\rm pc} \leq L_{\rm inj} \leq 10\,{\rm pc}$ brackets plausible turbulence scales in stellar cluster winds and superbubbles \citep{bykov2001,Morlino_2021}. Finally, we considered Kolmogorov, Kraichnan, and Bohm diffusion prescriptions, which span the standard range of physically motivated turbulence regimes.

\begin{figure*}[h]
\sidecaption
    \includegraphics[width = 12cm]{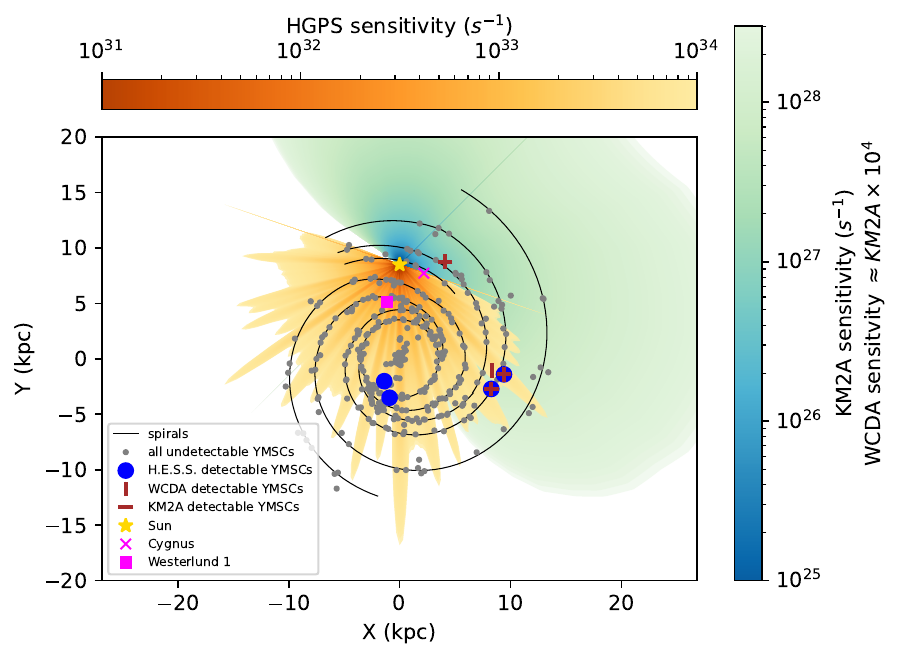}
		 \caption{Example of one realisation of a simulated population of Galactic YMSCs. The orange shaded region indicates the HGPS detectability range for point-like sources with varying luminosities. The edge of the region is for a luminosity of $10^{34}$ photons\,s$^{-1}$ (equivalent to a differential flux at 1\,TeV of $\sim 8 \times 10^{-11}\,{\rm TeV^{-1}\,cm^{-2}\,s^{-1}}$ at a distance of 1\,kpc). The blue-green shaded region shows the LHAASO detectability range for point-like sources with varying luminosities. The edge of the region is for luminosities of $1\times 10^{33}$ photons\,s$^{-1}$ and $3\times 10^{28}$ photons\,s$^{-1}$, corresponding to differential fluxes of $\sim 8 \times 10^{-12}\,{\rm TeV^{-1}\,cm^{-2}\,s^{-1}}$ at 3\,TeV and $\sim 3 \times 10^{-16}\,{\rm TeV^{-1}\,cm^{-2}\,s^{-1}}$ at 50\,TeV, respectively (at a distance of 1\,kpc) for the WCDA and KM2A. Grey dots mark simulated sources not detectable by any instrument, blue circles denote sources detectable only by H.E.S.S., vertical red lines identify those detectable by the WCDA, and horizontal red lines identify those detectable by the KM2A. In this realisation, 1.7\% of the YMSCs are detectable by at least one instrument. The Sun is marked by the yellow star, and the spiral arms of the Milky Way are shown in black. The parameter set used for this population is $\alpha = 4.5$, $\eta_{\rm CR} = 10^{-2.0}$, $\eta_{\rm b} = 10^{-2.0}$, $L_{\rm inj} = 1$\,pc, assuming the Kraichnan diffusion regime. The Cygnus bubble is shown by the pink cross, and the Westerlund 1 cluster is shown by the pink square.}

    	\label{fig:single_realisation}
\end{figure*}

\paragraph{}
We considered two different Galactic surveys, the HGPS and the First LHAASO Catalogue of Gamma-Ray Sources. For each source, using the gamma-ray luminosity, size, and position we were able to determine whether it would have been detected in the HGPS and the First LHAASO Catalogue of Gamma-Ray Sources. To account for the selection bias in the H.E.S.S. catalogue, we followed \citet{Steppa_2020}, and in particular we accounted for the fact that the HGPS sensitivity varies with position and source extent. For a simulated source to be detectable, it must exceed the threshold of $5\sigma$ above the background. This condition reads
\begin{equation}
\label{equation:detectability}
F_{\rm min} (\theta_{\rm source}) = \begin{cases}
F_{\rm min,0} \sqrt{\frac{\theta_{\rm source}^2 + \theta_{\rm PSF}^2}{\theta_{\rm PSF}^2}}, & \theta_{\rm source} \leq 1^{\circ} \\
\infty, & \theta_{\rm source} > 1^{\circ}
\end{cases} ,
\end{equation}
where $F_{\rm{min,0}}$ is the point-source sensitivity, $\theta_{\rm{source}}$ is the source extent (radius), and $\theta_{\rm PSF}$ is the size of the H.E.S.S. point-spread function. The limited field of view of H.E.S.S. combined with the applied technique of deriving background measurements means that sources of $\gtrsim 1^\circ$ in extent are not detectable. In Fig. \ref{fig:single_realisation} the HGPS detectability range for point-like sources with a luminosity of $10^{34} \text{photons s}^{-1}$ is displayed, demonstrating the large variations in sensitivity as a function of Galactic longitude. This variation in sensitivity shows why it is not sufficient to have a cut-off in the integrated flux to determine if a source is detectable and, thus, why we need to include the sensitivity of H.E.S.S. when testing each simulated source.

The First LHAASO Catalogue of Gamma-Ray Sources is split by detector (WCDA and KM2A); therefore, we needed a function for each detector to determine detectability. The sensitivity of each detector as a function of declination can be seen in \citet{LHAASO_catalogue}, Fig. 5. We considered the position, integrated flux (above 3 TeV for the WCDA and above 50 TeV for the KM2A), and extent of the source when determining its detectability. We account for the extent of the source in the same way as for the HGPS (see Equation~\ref{equation:detectability}), except that we set the maximum extent to the size of the largest object in the catalogue. 

\paragraph{}
Because many of the sources in the catalogues are unidentified, and at least some of those could be YMSCs~\citep{peron2024}, we created upper and lower limits for each of the catalogues to compare with. In adopting these limits, we took a conservative approach, since many YMSCs can be explained by a range of scenarios; this makes firm detections challenging to achieve. When confronting our populations with the HGPS, the lower limit is the YMSC Westerlund 1; \citet{westerlund_1} claimed that of the known objects within this region only the cluster can explain the majority of the emission. The upper limit is the lower limit plus the unidentified sources associated by position with known stellar clusters (5). These were taken from gamma-ray sources that overlap with a stellar cluster taken from \citet{peron2024}, Table 1. H.E.S.S. has detected emission from the region of Westerlund 2\citep{westerlund_2}; however, differently to Westerlund 1, the preferred emission scenario is not from the stellar cluster\citep{Asahina_2017}. When confronting our populations with the First LHAASO Catalogue of Gamma-Ray Sources, we applied the same procedure; the lower limit (for both WCDA and KM2A) is the Cygnus cocoon \citep{LHAASO_cygnus}. Although VHE emission from W43 has also been detected by LHAASO\citep{W43_LHAASO}---interpreted as support to the YMSC scenario---this makes it a similar case to the Cygnus cocoon emission where alternative scenarios can explain the emission. In this case we set a conservative lower limit to just one of these being from the YMSC, but we also monitored the effect on our results when including both and setting the lower limits for WCDA and KM2A to two. \citet{LHAASO_conference}, on behalf of the LHAASO collaboration, showed that in the First LHAASO Catalogue of Gamma-Ray Sources there is one detection of a massive cluster. The upper limits are the lower limit plus the sources associated by position with known stellar clusters: 12 for WCDA and 9 for KM2A. A summary of this can be seen in Table~\ref{table:Summary_detectable_table}. It should be noted that the upper limits for the LHAASO detectors are less important than the lower limit; this is because more populations are ruled out as they do not have enough YMSCs with sufficient energy to be detectable by LHAASO than due to them having too many LHAASO-detectable YMSCs, since when that happens they are usually already ruled out by the H.E.S.S. upper limit. Additionally, we have one more criterion: there should be at least two separate YMSCs detectable, since it is possible that one simulated YMSC can be detectable by all three instruments, but at least two were detected in the Galaxy. In Fig.~\ref{fig:single_realisation} a single realisation that matched all the above criteria is shown. However, it does not reproduce the exact locations of known YMSCs. This is because the model of the cluster is decoupled from the model determining the placement of the cluster, in other words, the population model does not include any information on the location of the cluster to determine the makeup of the stars in the cluster.

\begin{table}[h]
    \centering
    \caption{Summary table showing the number of gamma-ray-emitting YMSCs.}
    \begin{tabular}{|c|c|c|}
	\hline
	Catalogue & Detected & Associated \\
	\hline
	HGPS & 1 & 5 \\
	LHAASO WCDA & 1* & 12 \\
	LHAASO KM2A & 1* & 9 \\
	\hline

	\end{tabular}
    \tablefoot{This table shows the detected YMSCs and the number of gamma-ray sources associated to stellar clusters (i.e. overlapping with a stellar cluster), in the two catalogues, HGPS and LHAASO First Catalogue of Gamma-Ray Sources split by instrument. The upper limits are taken from \citet{peron2024}.\\
    \tablefoottext{*}{The detected YMSCs for the LHAASO instruments could in fact be two if both the Cygnus cocoon and W43 are considered YMSCs.}
    }
    \label{table:Summary_detectable_table}
\end{table}

\section{Results} \label{section:results}
We performed a systematic exploration of the parameter space: $4.0 \leq \alpha \leq 4.5$, $10^{-3} \leq \eta_{\rm CR} \leq 10^{-1}$, $10^{-3} \leq \eta_{\rm b} \leq 10^{-1}$, $0.1\,{\rm pc} \leq L_{\rm inj} \leq 10\,{\rm pc}$, and three diffusion regimes (Kolmogorov, Kraichnan, and Bohm). We constrained our simulated populations using the HGPS and the First LHAASO Catalogue of Gamma-Ray Sources. The results of this search are presented in Figs.~\ref{fig:Kolmogorov}, \ref{fig:Kraichnan}, and \ref{fig:Bohm}. Smaller values of $\alpha$, larger values of $\eta_{\rm CR}$ and $\eta_{\rm b}$, and a smaller $L_{\rm inj}$ increase both the integrated flux and the maximum energy to which particles can be accelerated, thereby enhancing the number of detectable YMSCs.

\paragraph{}
In Fig.~\ref{fig:Kolmogorov}, we show the results of the parameter search for populations following the Kolmogorov diffusion regime. Most parameter sets fail to reproduce the experimental data, with only eight yielding more than 50\% of realisations in agreement with the observations. The parameter set that best matches the data ($\alpha = 4.4$, $\eta_{\rm CR} = 10^{-2.5}$, $\eta_b = 10^{-3.0}$, and $L_{\rm inj} = 0.1$\,pc) results in 79\% of realisations lying within the observational limits.
In general, most simulated populations do not produce enough detectable sources, as the Kolmogorov diffusion regime does not accelerate particles to sufficiently high energies for the YMSCs to be observable. However, some parameter sets lead to too many detectable YMSCs to be compatible with the HGPS data (e.g. $\alpha = 4.0$, $\eta_{\rm CR} = 0.1$, $\eta_b = 0.1$, and $L_{\rm inj} = 0.1$\,pc).
Comparing the populations with data from individual instruments is less constraining than using the combined observational constraints. When confronted with only one instrument’s data, some parameter sets achieve 100\% of realisations within the limits. Overall, only two parameter sets provide a satisfactory match to the experimental data. The best matches are $\alpha = 4.5$, $\eta_{\rm CR} = 10^{-2.0}$, $\eta_b = 10^{-3.0}$, and $L_{\rm inj} = 0.1$\,pc, with 78\% of realisations lying within observational limits. When using Kolmogorov diffusion, none of the populations with $L_{\rm inj} = 10$\,pc are consistent with the observations, as they all produce too few detectable YMSCs.

\paragraph{}
Figure~\ref{fig:Kraichnan} illustrates the parameter exploration for populations adopting the Kraichnan diffusion regime. This regime allows for the acceleration of particles to higher energies than the Kolmogorov, and the best-fitting parameter set ($\alpha = 4.5$, $\eta_{\rm CR} = 10^{-2.0}$, $\eta_{\rm b} = 10^{-2.0}$, and $L_{\rm inj} = 1$\,pc) yields 79\% of realisations in agreement with all observational constraints. There are three parameter sets with at least 75\% of their realisations consistent with the experimental data; these span the ranges $4.2 \leq \alpha \leq 4.5$, $10^{-3} \leq \eta_{\rm CR} \leq 10^{-2}$, $10^{-2} \leq \eta_{\rm b} \leq 10^{-1.5}$, and $L_{\rm inj} = 1$\,pc. When using Kraichnan diffusion, there are 17 parameter sets that have more than 50\% of realisations in agreement with the data.

The most restrictive instrument in the Kraichnan case is H.E.S.S.; most parameter sets produce too many detectable YMSCs to remain compatible with the HGPS constraints. When confronting the populations with only one instrument’s data, we find parameter sets for which 100\% of the realisations lie within the limits. The LHAASO data are considerably less constraining, since most Kraichnan parameter sets produce YMSCs energetic enough to be detectable, and the upper limits of the LHAASO measurements do not significantly restrict the parameter space.

\paragraph{}
Figure~\ref{fig:Bohm} illustrates the exploration of the parameter space in the case of Bohm diffusion. Only one injection scale is considered, since $L_{\rm inj}$ does not affect the maximum particle energy in this regime. Bohm diffusion accelerates particles to higher energies than both the Kolmogorov and Kraichnan regimes. None of the parameter sets have any realisations that are in agreement with all the observational constraints. In the Bohm regime, populations are no longer rejected because they produce too few detectable YMSCs; instead, they are excluded almost exclusively for producing too many. The KM2A dataset provides the most stringent constraints; for all parameter sets, every realisation contains too many detectable YMSCs. The parameter set producing the least detectable YMSCs (although still too many) is $\alpha = 4.5$, $\eta_{\rm CR} = 0.001$, and $\eta_{\rm b} = 0.001$. When confronting the simulations with only a single instrument’s data, some parameter sets yield up to 99\% of their realisations within the observational limits. However, all parameter sets fail to reproduce the combined constraints when Bohm diffusion is assumed.

\paragraph{}
The overall best-matching parameter sets yield 79\% of realisations in agreement with the observational constraints. These correspond to $\alpha = 4.5$, $\eta_{\rm CR} = 10^{-2.0}$, $\eta_{\rm b} = 10^{-2.0}$, and $L_{\rm inj} = 1$\,pc when using the Kraichnan diffusion regime, and $\alpha = 4.4$, $\eta_{\rm CR} = 10^{-2.5}$, $\eta_{\rm b} = 10^{-3.0}$, and $L_{\rm inj} = 0.1$\,pc using the Kolmogorov regime. We find that combining all available observational data significantly reduces the acceptable parameter space, since the different instruments probe distinct energy ranges. The LHAASO data constrain the populations by requiring at least one simulated source to be detectable above 50\,TeV, while the HGPS data constrain the number of sources emitting above 1\,TeV. If we make the assumption that W43 is also a YMSC emitting TeV gamma rays, i.e. changing the lower limit to two, we find that all our percentages of realisations in agreement with the data drop around 5-10\%; for example, our best parameter sets have between 73\% and 66\% of realisations in agreement with the observations when using Kolmogorov diffusion and between 67\% and 71\% when using Kraichnan diffusion. Overall we found 14 parameter sets with at least 50\% of realisations in agreement with the data in this case. Another effect of this is that it rules out more of the parameter space around the regions where there are too few detections in Figs.~\ref{fig:Kolmogorov} and \ref{fig:Kraichnan}.

We also find that small changes in our assumptions can lead to large shifts in the preferred regions of parameter space. For example, using an older catalogue of YMSCs to infer their Galactic distributions results in far fewer estimated clusters than obtained here, thereby requiring YMSCs to be more efficient particle accelerators to account for the observed emission. Understanding the underlying Galactic distribution of YMSCs is therefore crucial for determining their role in high-energy particle acceleration. Our results are significantly more stringent than those obtained in our preliminary approach \citep{ICRC_proceedings}. This difference arises from the updated observational inputs for the cluster age and mass distributions used in the present model. In our previous work, these inputs led us to expect only $\sim 15$ YMSCs in our simulated populations, whereas the revised distributions yielded $\sim 300$. This highlights the critical role of observational inputs and their substantial impact on the resulting predictions.

\section{Conclusions} \label{section:conclusion}

Relying on a Monte Carlo approach, we simulated populations of the Galactic YMSCs in the TeV range (built on a physically motivated model of the gamma-ray emission from accelerated protons at YMSC shocks) and compared the results of the simulated populations with available data from the systematic survey of the Galactic plane performed by H.E.S.S. and of the First LHAASO Catalogue of Gamma-Ray Sources. We performed a systematic exploration of the parameter space defined by $\alpha$, the power-law spectral index of the accelerated particles at the YMSC shocks; $\eta_{\rm CR}$, the fraction of the wind luminosity converted into CRs; and $\eta_{\rm b}$, the fraction of the wind luminosity converted into a turbulent magnetic field; and we considered different prescriptions for the description of the maximum energy of accelerated particles, $E_{\rm max}$, and different prescriptions for the spatial distribution of sources. 
Our conclusions are listed below.
\begin{enumerate}

\item Despite the limitations of our approach, the systematic exploration of the parameter space allowed us to identify the region of the parameter space producing the most realisations within the limits of the HGPS sample and the First LHAASO Catalogue of Gamma-Ray Sources. In addition to the requirement concerning the total number of sources detectable by each instrument, including the total number detected in the population by any instrument, we can further reduce the allowed parameter space. Two possible solutions leading to $79$\% of realisations in the combined limits of the HGPS and the First LHAASO Catalogue of Gamma-Ray Sources are $\alpha = 4.5$, $\eta_{\rm CR} = 10^{-2.0}$, $\eta_{\rm b} = 10^{-2.0}$, and $L_{\rm inj} = 1$\,pc, using the Kraichnan diffusion regime; and $\alpha = 4.4$, $\eta_{\rm CR} = 10^{-2.5}$, $\eta_{\rm b} = 10^{-3.0}$, and $L_{\rm inj} = 0.1$\,pc, using the Kolmogorov regime. These solutions correspond to a rather steep spectrum of accelerated particles at the sources that are typically at the limit of or steeper than the $4.1-4.4$ range usually expected~\citep{strong2007,amato2014,cristofari2021}.

\item There are five sets of parameters with at least 75\% of realisations in agreement with the combined limits of the HGPS and the First LHAASO Catalogue of Gamma-Ray Sources. The two mentioned previously and the other three have the following parameters: $\alpha = 4.2, \: 4.4, \: 4.5$; $\eta_{\rm CR} = 10^{-3.0}, \: 10^{-2.5}, \: 10^{-2.0}$; and $\eta_{\rm b} = 10^{-2.0}, \: 10^{-1.5}, \: 10^{-3.0}$; respectively. The first two assume Kraichnan diffusion with $L_{\rm inj} = 1$ pc, and the third one assumes Kolmogorov diffusion with $L_{\rm inj} = 0.1$ pc. There are many more parameter sets with at least 50\% of realisations in agreement with observations.

\item Overall, we find that most of the parameter space of our model does not allow us to satisfactorily account for the observational data.
In the case of Bohm diffusion, no parameter sets reproduce the observations. This may be seen as an additional argument for the tension faced by YMSCs regarding their potential contribution to the CR spectrum (see e.g. the problem of the excess of grammage at $\sim 10$~TeV\citep{blasi2025}).

\item When considering only the limits from a single experiment (e.g. only the HGPS or only LHAASO), several sets of parameters are found to be in agreement with the data (>90 \% in agreement). This clearly shows the importance of combining the information from the two experiments.

\item We find that our results are sensitive to our input parameters, and it is important to achieve the best observational results to make solid conclusions about the particle acceleration in YMSCs.

\end{enumerate}

If it turns out that there should be more YMSCs than we estimated in this work, the favoured part of the parameter space would be the part producing less gamma-ray emission; i.e. larger $\alpha$, lower $\eta_{\rm CR}$, lower $\eta_{\rm b}$, and larger $L_{\rm inj}$, and vice versa if there are fewer YMSCs than we estimated. The same goes for changing the physics of the model; if one includes something that produces more gamma-ray emission the parts of the parameter space producing less emission will be favoured.

It is possible that including the SNR shocks in the modelling of the YMSCs could produce particle acceleration up to higher energies than our current model and produce more detectable simulated YMSCs~\citep{vieu2023}. An increased number of detections will be pushed even further with next-generation observatories such as CTAO~ \citep{cristofari2017,scienceCTA,CTA_galactic} or the LHAASO wide field air Cherenkov telescopes array~ \citep{LHAASO_IACT}. Assuming that our best performing population with the Kraichnan diffusion regime is realistic, we expect that CTAO will detect $\sim 17$ and that LHAASO (in twice the length of time of the first catalogue) will detect $\sim 3$ with both the WCDA and KM2A detectors. A firm identification of the YMSCs would greatly constrain our parameter space, as it would reduce the range of populations required to explain the data.

 \begin{acknowledgements}  
 This work has been funded by the Deutsche Forschungsgemeinschaft (DFG, German Research Foundation) with the grant 500120112. PC acknowledges support from the PSL Starting Grant GALAPAGOS.
 \end{acknowledgements}

\bibliography{aa58623-25}

\begin{appendix}
\onecolumn
\section{Plots showing the results of the systematic exploration of the parameter space}
\begin{figure*}[ht!]
    \centering
    \includegraphics[width = 0.94\linewidth]{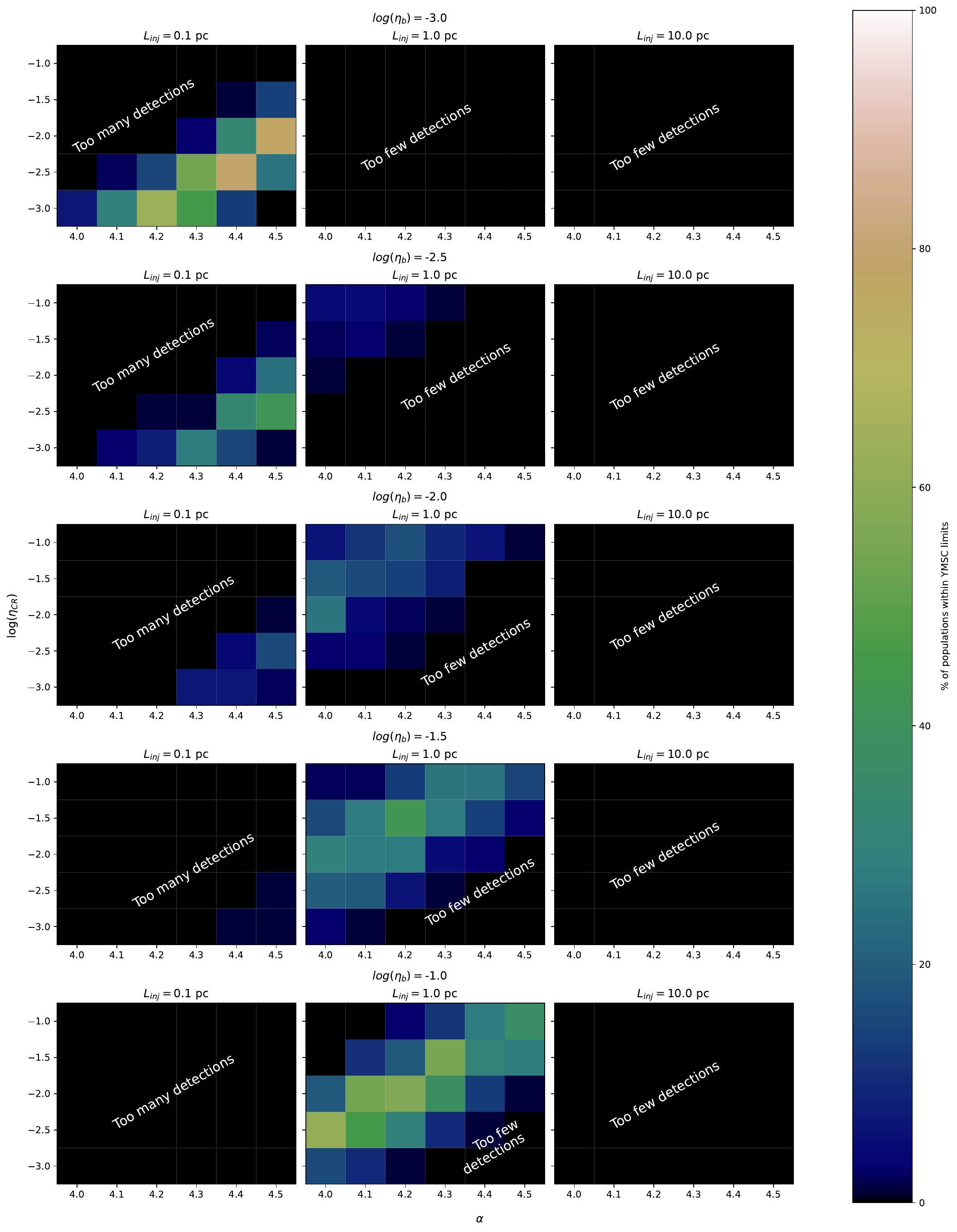}
    \caption{2D histograms showing the percentage of realisations that agree with the observational constraints. The $y$-axis is the fraction of wind luminosity converted into CRs, and the $x$-axis is the spectral index. Columns correspond to different injection scales, and rows indicate the fraction of wind luminosity converted into magnetic field. The Kolmogorov diffusion regime is assumed.
}
    \label{fig:Kolmogorov}
\end{figure*}

\begin{figure*}
    \centering
    \includegraphics[width = \linewidth]{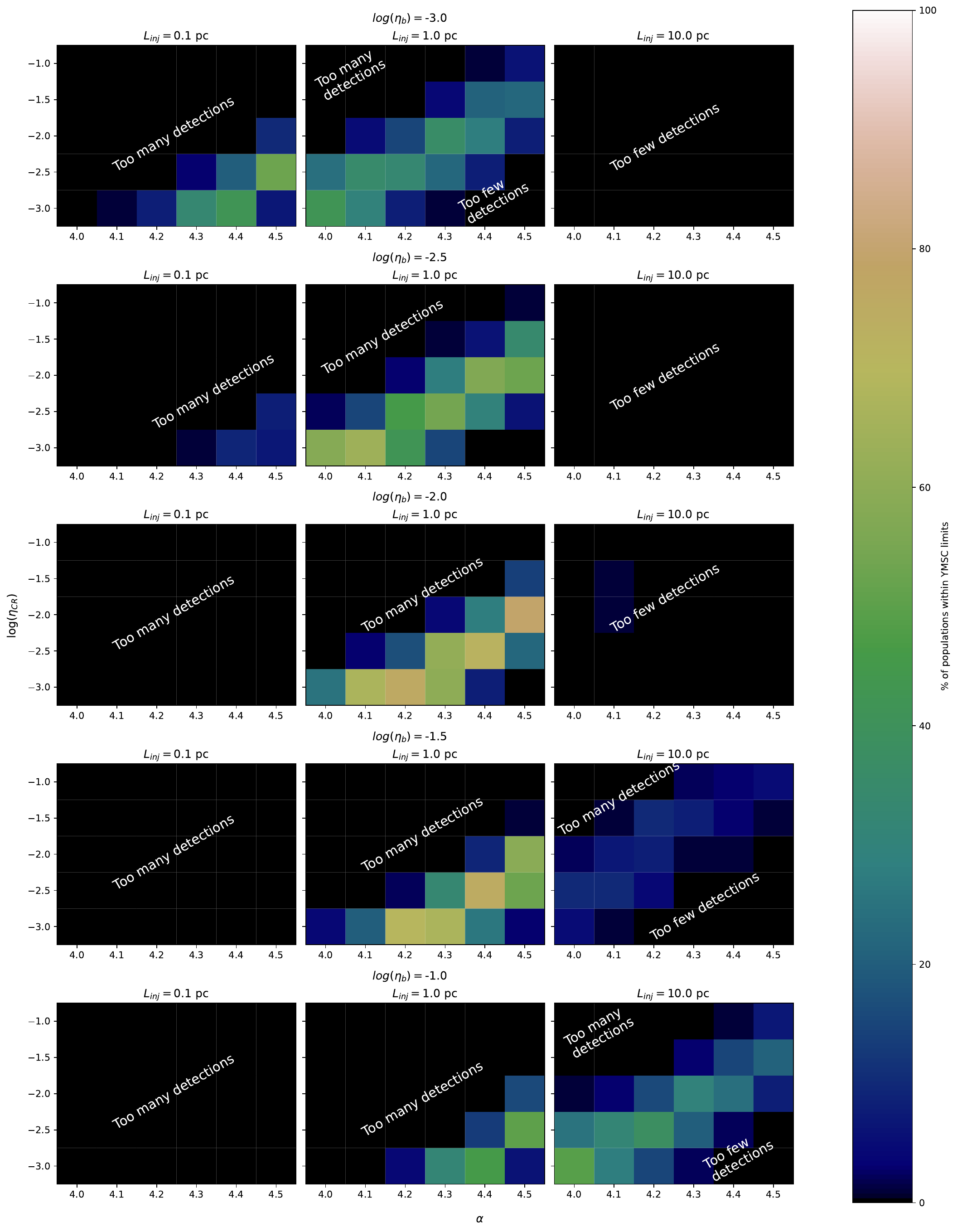}
    \caption{Same as Fig.~\ref{fig:Kolmogorov} with Kraichnan diffusion.}
    \label{fig:Kraichnan}
\end{figure*}

\begin{figure*}
    \centering
    \includegraphics[width = \linewidth]{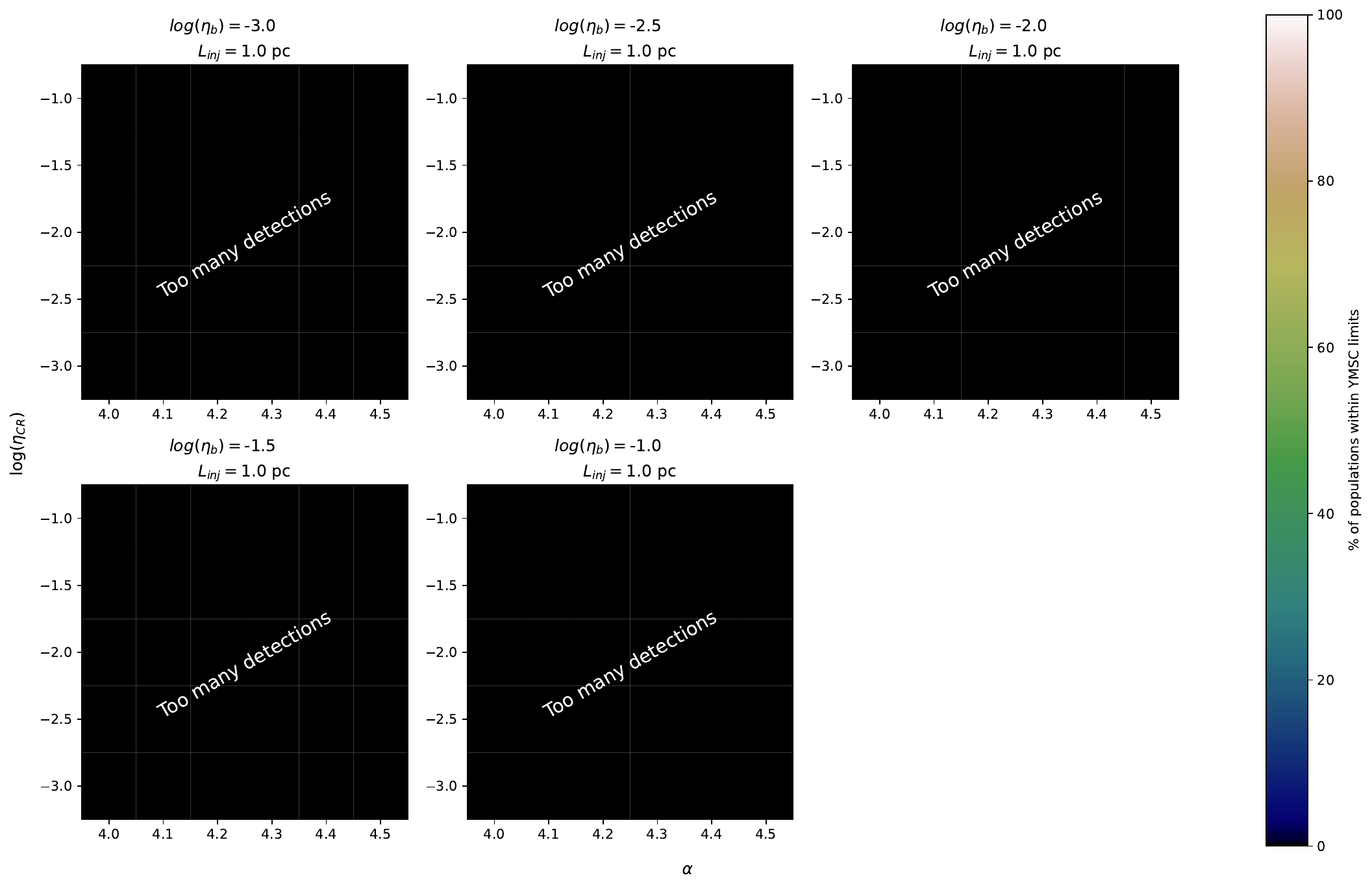}
    \caption{Same as Fig.~\ref{fig:Kolmogorov} with Bohm diffusion and only showing one turbulence injection scale since it does not change anything.}
    \label{fig:Bohm}
\end{figure*}

\end{appendix}
\end{document}